\documentclass[fleqn,twoside]{article}
\usepackage[headings]{espcrc2}
\usepackage{graphicx}
\usepackage{amsmath}
\usepackage{amssymb}
\usepackage{amsthm}
\usepackage{algorithmic}
\readRCS
$Id: espcrc2.tex,v 1.2 2004/02/24 11:22:11 spepping Exp $
\usepackage{graphicx}
\usepackage[figuresright]{rotating}

\newcommand{\Gr}{Gr\"obner }
\newcommand{\cJ}{\cal{J}}

\newcommand{\Z}{\mathbb{Z}}
\newcommand{\K}{\mathbb{K}}

\newcommand{\R}{\mathbb{R}}

\newcommand{\lm}{\mathop{\mathrm{lm}}\nolimits}
\newcommand{\lt}{\mathop{\mathrm{lt}}\nolimits}
\newcommand{\lc}{\mathop{\mathrm{lc}}\nolimits}

\newcommand{\Card}{\mathop{\mathrm{Card}}\nolimits}

\newcommand{\Id}{\mathop{\mathrm{Id}}\nolimits}

\newenvironment{algorithm}[1]{
  \begin{center}
    {\bf Algorithm: #1}\\*
     \begin{tabular}{|p{70mm}|} \hline
} {
 \\ \hline
 \end{tabular}
 \end{center}
}

\title{On Computation of \Gr Bases for Linear Difference Systems}

\author{Vladimir P.Gerdt\address[MCSD]{Laboratory of Information Technologies,
        Joint Institute for Nuclear Research,
        141980 Dubna, Russia}\thanks{gerdt@jinr.ru}}

\runtitle{On Computation of \Gr Bases for Linear Difference Systems}
\runauthor{V.P.Gerdt}

\begin{document}

\begin{abstract}
In this paper we present an algorithm for computing \Gr bases of
linear ideals in a difference polynomial ring
over a ground difference field. The input difference
polynomials generating the ideal are also assumed to be linear.
The algorithm is an adaptation to difference ideals of our
polynomial algorithm based on Janet-like reductions.
\vspace{1pc}
\end{abstract}

\maketitle

\section{INTRODUCTION}

Being invented 40 years ago by Buchberger~\cite{Buch65} for algorithmic solving of the
membership problem in the theory of polynomial ideals, the \Gr bases method has become a
powerful universal algorithmic tool for solving various mathematical problems arising in
science and engineering.

Though overwhelming majority of the \Gr bases applications is still found in commutative
polynomial algebra, over the last decade a substantial progress has also
been achieved in application
of \Gr bases to noncommutative polynomial algebra, to algebra of differential operators
and to linear partial differential equations (see, for example, book~\cite{GBA}). As to the
difference algebra, i.e. algebra of difference polynomials, in spite of its conceptual
algorithmic similarity to differential algebra, only a few efforts have been done to
extend the theory of \Gr bases to difference algebra and to exploit their algorithmic
power~\cite{Ch98,Mich'99}.

Recently, two promising applications of difference \Gr bases were revealed:
generation of difference schemes for numerical solving of PDEs~\cite{MB01,GBM05} and
reduction of multiloop Feynman integrals to the minimal set of basis integrals~\cite{G04}.

In this note we describe an algorithm (Sect.4) for constructing \Gr bases for linear difference
systems that is an adaptation of our polynomial algorithm~\cite{GB05} to linear difference ideals.
We construct a \Gr basis in its Janet-like form (Sect.3), since this approach has shown its
computational efficiency in the polynomial case~\cite{GB05,InvAlg}. We briefly outline
these efficiency issues in Sect.5. The difference form of the algorithm exploits some basic notions and
concepts of difference algebra (Sect.2) as well as the definition of Janet-like \Gr bases and Janet-like
reductions together with the algorithmic characterization of Janet-like bases (Sect.3). We conclude in
Sect.6.

\section{ELEMENTS OF DIFFERENCE ALGEBRA}

Let $\{y^1,\ldots,y^m\}$ be the set of {\em indeterminates}, for example, functions
of $n-$variables $\{x_1,\ldots,x_n\}$, and $\theta_1,\ldots,\theta_n$ be the set of mutually commuting
{\em difference operators (differences)}, e.g.,
$\theta_i\circ y^j=y^j(x_1,\ldots,x_i+1,\ldots,x_n).$

A {\em difference ring $R$ with differences
$\theta_1,\ldots,\theta_n$} is a commutative ring $R$ such that
$\forall f,g\in R,\ 1\leq i,j\leq n$
$$
\begin{array}{l}
\theta_i\theta_j=\theta_j\theta_i,\ \theta_i \circ (f+g)=\theta_i \circ f+\theta_i \circ g,\\
\theta_i \circ (f\,g)=( \theta_i \circ f) (\theta_i \circ g)\,.
\end{array}
$$
Similarly, one defines a {\em difference field}.

Let $\K$ be a difference field, and $\R:=\K\{y^1,\ldots,y^m\}$ be
the difference ring of polynomials over $\K$ in variables
$$\{\ \theta^\mu \circ y^k\ \mid \mu\in \Z^n_{\geq 0},\,k=1,\ldots,m\ \}\,.$$
Hereafter, we denote by $\R_L$ the set of linear polynomials in $\R$ and use the
notations:
$$
\begin{array}{l}
\Theta=\{\,\theta^\mu\ \mid\ \mu\in \Z^n_{\geq 0}\ \},\
\deg_i(\theta^\mu\circ y^k)=\mu_i, \label{theta} \\[0.1cm]
\deg(\theta^\mu \circ
y^k)=|\mu|=\sum_{i=1}^n \mu_i\,.
\end{array}
$$

A {\em difference ideal} is an ideal $I \subseteq \R$ closed under the action of any
operator from $\Theta$. If $F:=\{f_1,\ldots,f_k\}\subset \R$ is a
finite set, then the smallest difference ideal containing $F$
will be denoted by $\Id(F)$. If for an ideal $I$ there is $F\subset \R_L$ such that $I=\Id(F)$,
then $I$ is a {\em linear difference ideal}.

A total ordering $\succ$ on the set of
$\theta^\mu \circ y^{\,j}$ is a {\em ranking} if $\forall \,i,j,k,\mu,\nu$ the following
hold:
$$
\begin{array}{l}
\theta_{i} {\theta^\mu \circ y^{\,j}} \succ {\theta^\mu}\circ y^{\,j}\,,\\
\theta^\mu \circ y^{\,j} \succ \theta^\nu \circ y^k \iff
{\theta_i}  {\theta^\mu \circ y^{\,j}}
   \succ {\theta_i}{\theta^\nu} \circ y^k\,.
\end{array}
$$

If $\mu \succ \nu \Longrightarrow {\theta^\mu \circ y^{\,j}} \succ {\theta^\nu} \circ y^k$
the ranking is {\em orderly}. If $j \succ k \Longrightarrow {\theta^\mu} \circ y^{\,j} \succ {\theta^\nu} \circ y^k$
the ranking is {\em elimination}.

Given a ranking $\succ$, a linear polynomial $f\in \R_L\setminus \{0\}$
has the {\em leading term}
$a\,\vartheta \circ y^j$, $\vartheta\in \Theta$, where $\vartheta \circ y^j$ is maximal
w.r.t. $\succ$ among all $\theta^\mu \circ y^k$ which appear with nonzero
coefficient in $f$.
$\lc(f):=a\in \K\setminus \{0\}$ is the {\em leading coefficient}
and $\lm(f):=\vartheta \circ y^{\,j}$
is the {\em leading monomial}.

A ranking acts in $\R_L$ as a {\em monomial order}.
If $F \subseteq \R_L \setminus \{ 0 \}$, $\lm(F)$ will denote the set of the leading monomials and $\lm_j(F)$ will denote
its subset for indeterminate $y^{\,j}$.
Thus, $$\lm(F)=\cup_{j=1}^m \lm_j(F)\,.$$

\section{JANET-LIKE GR\"{O}BNER BASES}

Given a nonzero linear difference ideal $I=\Id(G)$ and a ranking $\succ$,
the ideal generating set
$G=\{g_1,\ldots,g_s\}\subset \R_L$ is a {\em \Gr basis}~\cite{GBA,Mich'99}
of $I$ if $\forall f\in I\cap \R_L\setminus \{0\}$:
\begin{equation}
\exists \, g\in G, \theta \in \Theta\  :\
\lm(f)=\theta \circ \lm(g)\,. \label{GB}
\end{equation}
It follows that $f\in I\setminus \{0\}$ {\em is reducible modulo $G$}:
$$ f \xrightarrow[g]{} f':=f-\lc(f)\,\theta \circ (g/\lc(g)),\quad f'\in I\,. $$
If $f'\neq 0$, then it is again reducible modulo $G$, and, by repeating the reduction, in finitely
many steps we obtain
$$\quad f \xrightarrow[G]{}0\,.$$
Similarly, a nonzero polynomial $h\in \R_L$, whose terms are reducible (if any) modulo
a set $F \subset \R_L$,
can be reduced to an irreducible
polynomial $\bar{h}$, which is said to be in the {\em normal form modulo $F$}
(denotation: $\bar{h}=NF(h,F)$).

In our algorithmic construction of \Gr bases we shall use a restricted
set of reductions called {\em Janet-like} (cf.~\cite{GB05}) and defined
as follows.

For a finite set $F \subseteq \R_L$ and a ranking $\succ$, we partition
every $\lm_k(F)$ into groups
labeled by $d_0,\ldots,d_i\in \Z_{\geq 0}$,\ $(0\leq i\leq n)$. Here $[0]_k:=\lm_k(F)$
and for $i>0$ the group $[d_0,\ldots,d_i]_k$ is defined as
$$
\{u\in \lm_k(F) \mid
d_0=0,d_j=\deg_j(u),1\leq j\leq i \}.
$$
Denote by $h_i(u,\lm_k(F))$ the nonnegative integer
$$\max\{\deg_i(v) \mid u,v\in [d_0,...,d_{i-1}]_k\}-\deg_i(u).$$
If $h_i(u,\lm_k(F))>0$, then $\theta_i^{s_i}$ such that
$$
\begin{array}{l}
s_i:=\min\{\deg_i(v)-\deg_i(u) \mid \\
u,v\in [d_0,...,d_{i-1}]_k,\, \deg_i(v)>\deg_i(u)\}
\end{array}
$$
is called a {\em difference power} for $f\in F$ with $\lm(f)=u$.

Let $DP(f,F)$ be the set of difference powers for $f\in F$, and
${\cJ}(f,F):=\Theta \setminus \bar{\Theta}$ be the subset of $\Theta$
with
\[
\bar{\Theta}:=\{\theta^\mu \mid \exists \, \theta^\nu \in DP(f,F)\ :\
\mu-\nu\in \Z^n_{\geq 0} \}.
\]
A \Gr basis $G$ of $I=\Id(G)$ is called Janet-like~\cite{GB05} if
$\forall f\in I\cap \R_L\setminus \{0\}$:
\begin{equation}
\exists \, g\in G, \theta
\in {\cJ}(g,G)\  :\ \lm(f)=\theta \circ
  \lm(g)\,. \label{JLGB}
\end{equation}
This implies ${\cJ}-$reductions and the ${\cJ}-$normal form $NF_{\cJ}(f,F)$. It is clear
that condition~(\ref{JLGB}) implies (\ref{GB}). Note, however, that the converse is generally not
true. Therefore, not every \Gr basis is Janet-like.

The properties of a Janet-like basis are very similar to those of a Janet basis~\cite{InvAlg}, but
the former is generally more compact than the latter. More preciously, let
$GB$ be a reduced
\Gr basis~\cite{GBA}, $JB$ be a minimal Janet basis, and $JLB$ be a minimal
Janet-like basis of the same ideal
for the same ranking. Then their cardinalities satisfy % the inequality
\begin{equation}
\Card(GB)\leq \Card(JLB)\leq \Card(JB) \label{cardinalities},
\end{equation}
where $\Card$ abbreviates {\em cardinality}, that is, the number of elements.

Whereas the algorithmic characterization of a \Gr basis is zero redundancy of all its $S-$
polynomials~\cite{Buch65,GBA}, the algorithmic characterization of a Janet-like basis $G$ is the
following condition~(cf.~\cite{GB05}):
\begin{equation}
\forall g\in G,\ \vartheta\in DP(g,G):\,NF_{\cJ}(\vartheta \circ g,G)=0\,. \label{alg_char}
\end{equation}
This condition is at the root of the algorithmic construction of Janet-like bases as described in the next section.

\section{ALGORITHM}

\begin{algorithm}{Janet-like \Gr Basis($F,\succ$)\label{JLB}}
\begin{algorithmic}[1]
\INPUT $F \subseteq \R_{L}\setminus \{0\}$, a finite set;\ $\succ$, a ranking
\OUTPUT $G$, a Janet-like basis of $\Id(F)$
\STATE {\bf choose} $f\in F$ with the lowest $\lm(f)\qquad$ w.r.t. $\succ$
\STATE $G:=\{f\}$
\STATE $Q:=F\setminus G$
\DOWHILE
  \STATE $h:=0$
  \WHILE{$Q\neq \emptyset$\ and $h=0$}
    \STATE {\bf choose} $p\in Q$ with the lowest $\lm(p)$ w.r.t. $\succ$
    \STATE $Q:=Q\setminus \{p\}$
    \STATE $h:={\bf Normal\ Form}(p,G,\succ)$
  \ENDWHILE
  \IF{$h\neq 0$}
    \FORALL{$g\in G$ such that $\lm(g)=\theta^\mu \circ \lm(h),\ |\mu|>0$}
      \STATE $Q:=Q\cup \{g\}$; \ $G:=G\setminus \{g\}$
    \ENDFOR
    \STATE $G:=G\cup \{ h \}$
    \STATE $Q:=Q\cup \{\,\theta^\beta\circ g \mid g\in G,\
                                    \theta^\beta\in DP(g,G)\,\}$
 \ENDIF
\ENDDO{$Q \neq \emptyset$}
\RETURN $G$
\end{algorithmic}
\end{algorithm}

This algorithm is an adaptation of the polynomial version~\cite{GB05} to linear difference
ideals. It outputs a minimal Janet-like \Gr basis which (if monic, that is, normalized by division
of each polynomial by its leading coefficient) is uniquely defined by the input set $F$ and ranking
$\succ$. Correctness and termination of the algorithm follow from the proof given in~\cite{GB05};
in so doing the displacement of some elements of the intermediate sets $G$ into $Q$ at step 13 provides
minimality of the output basis. The algorithm terminates when the set $Q$ becomes empty in accordance
with~(\ref{alg_char}).

The subalgorithm {\bf Normal\ Form}$(p,G,\succ)$ performs the Janet-like reductions (Sect.3) of the input difference
polynomial $p$ modulo the set $G$ and outputs the Janet-like normal form of
$p$. As long as the intermediate
difference polynomial $h$ has a term Janet-like reducible modulo $G$, the elementary reduction of this term is
done at step 4. As usually in the \Gr bases techniques~\cite{GBA}, the reduction is terminated in finitely many
steps due to the properties of the ranking (Sect.2).
\begin{algorithm}{Normal Form$(p,G,\succ)$}
\begin{algorithmic}[1]
\INPUT $p\in \R_{L}\setminus \{0\}$, a polynomial; $G\subset \R_{L}\setminus \{0\}$, a finite set;
       $\succ$, a ranking
\OUTPUT $h=NF_{\cJ}(p,G)$, the ${\cJ}-$normal form of $p$ modulo $G$
      \STATE $h:=p$
      \WHILE{$h\neq 0$ {\bf and} $h$ has a monomial $u$ with coefficient
              $b\in \K$ ${\cJ}-$reducible modulo $G$}
            \STATE {\bf take} $g\in G$ such that $u=\theta^{\gamma} \circ \lm(g)$
             with $\theta^{\gamma}\in {\cJ}(\lm(g),\lm(G))$
            \STATE $h:=h/b - \theta^{\gamma}\circ (g/\lt(g))$
      \ENDWHILE
  \RETURN $h$
\end{algorithmic}
\end{algorithm}
An improved version of the above algorithm can easily be derived from
the one for the involutive algorithm~\cite{InvAlg} if one replaces the input involutive division
by a Janet-like monomial division~\cite{GB05} and then translates the algorithm into linear difference algebra.
In particular, the improved version includes Buchberger's criteria adjusted to Janet-like
division and avoids the repeated prolongations $\theta^\beta \circ g$ at step 16 of the algorithm.

\section{COMPUTATIONAL ASPECTS}

The polynomial version of algorithm {\bf Janet-like \Gr Basis} implemented in its improved form in
C++~\cite{GB05} has disclosed its high computational efficiency for the standard set of benchmarks\footnote{See Web page
{\tt http://invo.jinr.ru}.}. If one compares this algorithm with the involutive one~\cite{InvAlg} specialized
in Janet division, then all the computational merits of the latter algorithm are retained, namely:
\begin{itemize}
\item Automatic avoidance of some useless reductions.
\item Weakened role of the criteria: even without applying any criteria the algorithm is reasonably fast.
By contrast, Buchberger's algorithm without applying the criteria becomes unpractical even for rather
small problems.
\item Smooth growth of intermediate coefficients.
\item Fast search of a polynomial reductor which provides an elementary Janet-like reduction of the given term.
It should be
noted that as well as in the involutive algorithm such a reductor, if it exists, is unique. The fast search is based on the
special data structures called Janet trees~\cite{InvAlg}.
\item Natural and effective parallelism.
\end{itemize}
Though one needs intensive benchmarking for linear difference systems, we have solid
grounds to believe that the above listed computational merits hold also for the difference case.

As this takes place, computation of a Janet-like basis is more efficient than
computation of a Janet basis by the involutive algorithm~\cite{InvAlg}. The
inequality~(\ref{cardinalities}) for monic bases is a consequence of the inclusion~\cite{GB05}:
\begin{equation}
GB\subseteq JLB\subseteq JB\,. \label{subsets}
\end{equation}
There are many systems for which the cardinality of a Janet-like basis is much closer to that of the reduced \Gr basis
than the cardinality of a Janet basis. Certain binomial ideals called toric form an important class of such problems.
Toric ideals
arise in a number of problems of algebraic geometry and closely related to integer programming. For this class
of ideals the cardinality of Janet bases is typically much larger than that of reduced \Gr bases~\cite{GB05}.
For illustrative purposes consider a difference analogue of the simple toric ideal~\cite{GB05,BLR99} generated in the
ring of difference operators by the following set:
$$\{\ \theta_x^7-\theta_y^2\theta_z, \theta_x^4\theta_w-\theta_y^3, \theta_x^3\theta_y-\theta_z\theta_w\ \}\,.$$
The reduced \Gr basis for the degree-reverse-lexicographic ranking with $\theta_x\succ \theta_y\succ
\theta_z\succ \theta_w$ is given by
$$ \{\ \theta_x^7-\theta_y^2\theta_z, \theta_x^4\theta_w-\theta_y^3, \theta_x^3\theta_y-\theta_z\theta_w,
\theta_y^4-\theta_x\theta_z\theta_w^2\ \}\,.$$
The Janet-like basis computed by the above algorithm contains one more element $\theta_x^4\theta_w-\theta_y^3$
whereas the Janet basis adds another six extra elements to the Janet-like basis~\cite{GB05}.

The presence of extra elements in a Janet basis in comparison with a Janet-like basis is obtained
because of certain additional
algebraic operations. That is why the computation of a Janet-like basis is
more efficient than the computation of a Janet basis.
Both bases, however, contain the reduced \Gr basis as the
internally fixed~\cite{InvAlg} subset of the output
basis\footnote{In the improved
versions of the algorithms.}.
Hence, having any of the bases computed, the reduced \Gr basis is easily extracted without any extra
computational costs.

\section{CONCLUSION}

The above presented algorithm
is implemented, in its improved form, as a Maple
package~\cite{GR-ACAT}, and already applied to generation of difference schemes for PDEs and to reduction
of some loop Feynman integrals (see some examples in~\cite{GR-ACAT}). The last problem
for more than 3 internal lines with masses is computationally hard for the current version of the package.

One reason for this is that the Maple implementation does not support Janet trees since Maple does not provide
efficient data structures for trees.

Another reason is that in the improved version of the algorithm there is
still some freedom in the selection strategy for elements in $Q$ to be reduced modulo $G$. Though our
algorithms are much less sensitive to the selection strategy than Buchberger's algorithm, the running time
still depends substantially on the selection strategy: mainly because of dependence of the intermediate
coefficients growth on the selection strategy. To find a heuristically good selection strategy
one needs to do intensive benchmarking with difference systems. In turn, this requires an extensive data base
of various
benchmarks that, unlike polynomial benchmarks, up to now is missing for difference systems.

In addition to our further research on improvements of the Maple package, we are  going to implement
the difference algorithm in C++ as a module of the open source software GINV available on the Web site
{\tt http://invo.jinr.ru}.

The comparison of implementations of polynomial involutive algorithms for Janet bases in Maple and
in C++~\cite{Daniel} shows that the C++ code is of two or three order faster than its Maple
counterpart. Together with efficient parallelization of the algorithm this gives a real hope
for its practical applicability to problems of current interest in reduction of loop integrals.

\section{ACHNOWLEDGEMENTS}

This work was partially supported by grants 04-01-00784 and 05-02-17645 from the
Russian Foundation for Basic Research and by grant 2339.2003.2
from the Russian Ministry of Science and Education. The author thanks
the organizers of ACAT 2005 for waiving the registration fee.

\end{document}